\documentclass{aa}
\usepackage{graphicx,natbib}
\begin{document}
   \title{Search for star clusters close to the Galactic plane with DENIS}

   \author{C\'eline Reyl\'e \inst{1} \and Annie C. Robin \inst{1}}

   \institute{CNRS UMR6091, Observatoire de Besan\c{c}on, BP1615, 
    25010 Besan\c{c}on Cedex, France\\
    e-mail: celine@obs-besancon.f,annie.robin@obs-besancon.f\\
	}

   \offprints{C\'eline Reyl\'e}
   \date{Received ; accepted }

   \titlerunning{Search for star clusters with DENIS}

   \abstract{An automated search for star clusters close to the Galactic
plane ($|b| < 5^o$) was carried out on the Point Source Catalogue of the
DENIS survey. 44\% of the Galactic plane have been 
observed and calibrated. The method allowed to retrieve 22 known star
clusters and to identify two new ones, not published yet although previously 
presented in the 2MASS web site as embedded clusters in HII regions.
Extinction in the field and in front of the clusters are estimated using a
model of population synthesis. We present the method and give the properties of 
these clusters. 
   }

   \maketitle

\section{Introduction}
The highly concentrated dust in the plane of the Milky Way makes difficult 
the observation in the visible range of embedded objects and most distant 
objects hidden behind the Galactic plane. However, the extinction decreases 
at longer wavelengths: in the K-band, it is about ten times smaller than in
the visible. Thus the recent Near-Infrared surveys provide suitable data for
systematic search for new objects projected on the Galactic plane. 
Using the \textit{Two Micron All Sky Survey} 
\citep[hereafter 2MASS,][]{Skrutskie1997}
in the J (1.25 $\mu m$), H (1.65 $\mu m$) and K$_s$ (2.17 $\mu m$) bands, 
\cite{Dutra2000} carried out a systematic search for new star clusters in a
5$^o \times 5^o$ field centered close to the Galactic plane and listed 58
IR star clusters or candidates. They also investigated selected regions with
evidence of star formation and found 42 star clusters or candidates 
\citep{Dutra2001}. \cite{Hurt2000} serendipitously discovered two globular
clusters near the Galactic plane. \cite{Vauglin2001} undertook a search for
galaxies by eye on J and K$_s$-band images from the 
\textit{Deep-Near Infrared Survey of the Southern Sky} 
\citep[hereafter DENIS,][]{Epchtein1997,Epchtein1999} and 
found 13 star clusters (Rousseau, private communication).

Using the data in I (0.82 $\mu m$), J, and mostly K$_s$ bands from DENIS, 
we performed an 
automated search for star clusters in a $\pm$5$^o$ band around the Galactic 
plane.
In Sect.~\ref{method} we describe our method and detection criteria. The
basic properties of the new star clusters are given in 
Sect.~\ref{results}. 

\section{Method}
\label{method}
DENIS observations are strips 12$'$ wide in right ascension, and 30$^o$ 
spanned in declination. A strip is made of 180 individual 12$' \times 12'$ 
images, with 2$'$ overlaps between consecutive images. The images are 
processed
at the Paris Data Analysis Center, and point source catalogues are extracted.
Over the 3317 strips calibrated at that time, we selected the strips which 
contain sources with Galactic latitude $|b| < 5^o$, that is 1219 strips.
They all together cover 44\% of the entire Galactic plane, with Galactic
longitudes ranging between $-160^o$ and $40^o$ (82\% of the strips are 
calibrated within this range).

The two criteria used to search for star clusters are based on the density of
stars and the integrated flux in the K$_s$ band. This band allows to probe 
more extinguished regions than the $I$ and $J$ bands. Each image of the strip 
is divided in 2$' \times 2'$ frames. The 
number of stars $N$ and total flux $F$ in the K$_s$ band are computed in each 
frame, using the
corresponding calibrated catalogue. The mean $\bar{N}$ and $\bar{F}$, and the 
dispersion around the mean $\sigma_N$ and $\sigma_F$ of these two parameters 
are computed for the entire image. A 2$' \times 2'$
window is then displaced with 0.4$'$ steps in declination and right ascension 
on the image. If the density of stars $N$ and the total flux $F$ in the 
window are 4 $\times \sigma_N$ and 4 $\times \sigma_F$ above the mean 
$\bar{N}$ and $\bar{F}$, the program indicates the eventuality of a detection.
The threshold has been defined with the faintest globular cluster discovered
with 2MASS, GC01.

However, the star cluster GC01 is not easily 
distinguishable only from the catalogue, and these two criteria only are not
sufficient to perform an automatic detection. Most of the stars in the cluster
have J-K$_s$ $>$ 3.5, which is not surprising for a globular cluster dominated
by red giants in a high extinction region. With the condition
J-K$_s$ $>$ 3.5, the cluster can now be detected in an
automatic way. A difficulty remains in the determination of the J-K$_s$ limit
due to the unknown color of an eventual cluster and to the variable extinction
from frame to frame. We chose to consider the middle of the color range of 
the stars, $\mbox{J-K}_{\mbox{\scriptsize mid}}$, in the 12$' \times 12'$ image as a tentative 
limit. The efficiency of the detection being very dependent 
on the selection of the limit J-K$_{\mbox{\scriptsize mid}}$ considered, we performed the 4 
$\times \sigma_N$ and 4 $\times \sigma_F$ test 3 times, over stars redder 
than J-K$_{\mbox{\scriptsize mid}}$ - 0.5, J-K$_{\mbox{\scriptsize mid}}$, and J-K$_{\mbox{\scriptsize mid}}$ + 0.5.
In the case of GC01, the determined limit is J-K$_{\mbox{\scriptsize mid}}$ = 3. The 
detection program fails with the condition J-K$_{\mbox{\scriptsize mid}} >$ 2.5 or 3, but 
is successful for J-K$_{\mbox{\scriptsize mid}} >$ 3.5. The variation of the J-K$_s$ limit 
increases the efficiency of detection, but also allows to detect younger star 
clusters than this globular cluster.

\cite{Wilson1991} and \cite{Battinelli1991} described automatic 
procedures for identification of star associations. They are based on 
agglomerative algorithms. Whereas \cite{Wilson1991} determined a search 
radius, \cite{Battinelli1991} algorithm, called 
Path-Linkage-Criterion technique, assigns two stars to the same group if it is
possible to go from one to the other jumping from star to star in steps 
smaller than a determined scale-length. It allows to detect compact and
filamentary groups. Our method is not optimized to
identify associations of young stars. It uses a color criterion in order to
enhance the contrast with background stars, combined together with a density 
criterion and a total luminosity criterion. Applied to Near-Infrared data, our
technique allows to retrieve either old globular clusters 
or young star clusters embedded in HII regions.

\section{Results}
\label{results}

\begin{figure*}
\centering
\includegraphics[width=5.7cm,clip=]{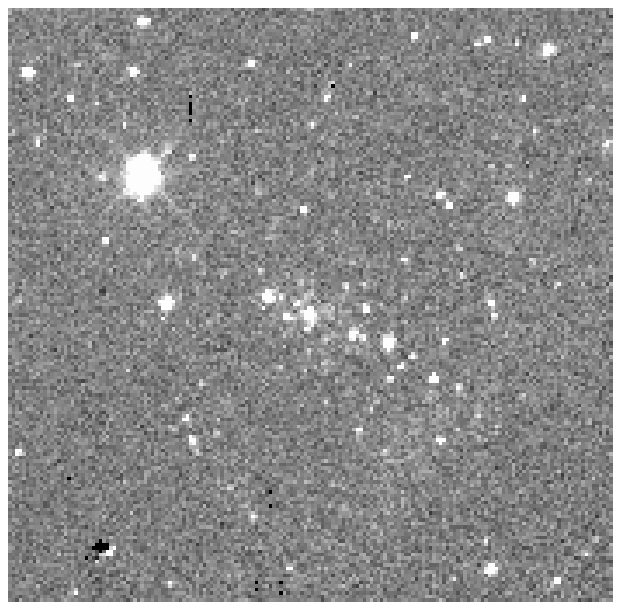}\hspace*{0.5cm}
\includegraphics[width=5.7cm,clip=]{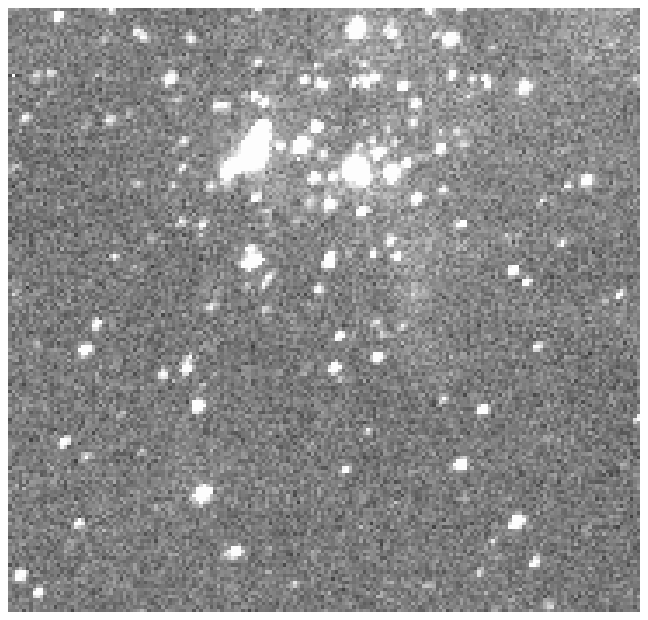}
   \caption{DENIS K$_s$-band images of the new star clusters. Left: Gum 25 cluster. Right: W40 
cluster. The size of the images is 5.8$' \times 5.8'$. 
Top = North, left = East.}
   \label{fig1}
\end{figure*}

\begin{figure*}
\hspace*{.8cm} (a) \hspace*{5.15cm} (b) \hspace*{5.15cm} (c) \vspace*{-.75cm} \\
\includegraphics[width=5.1cm,clip=,angle=-90,bb=97.5 60 572 637]{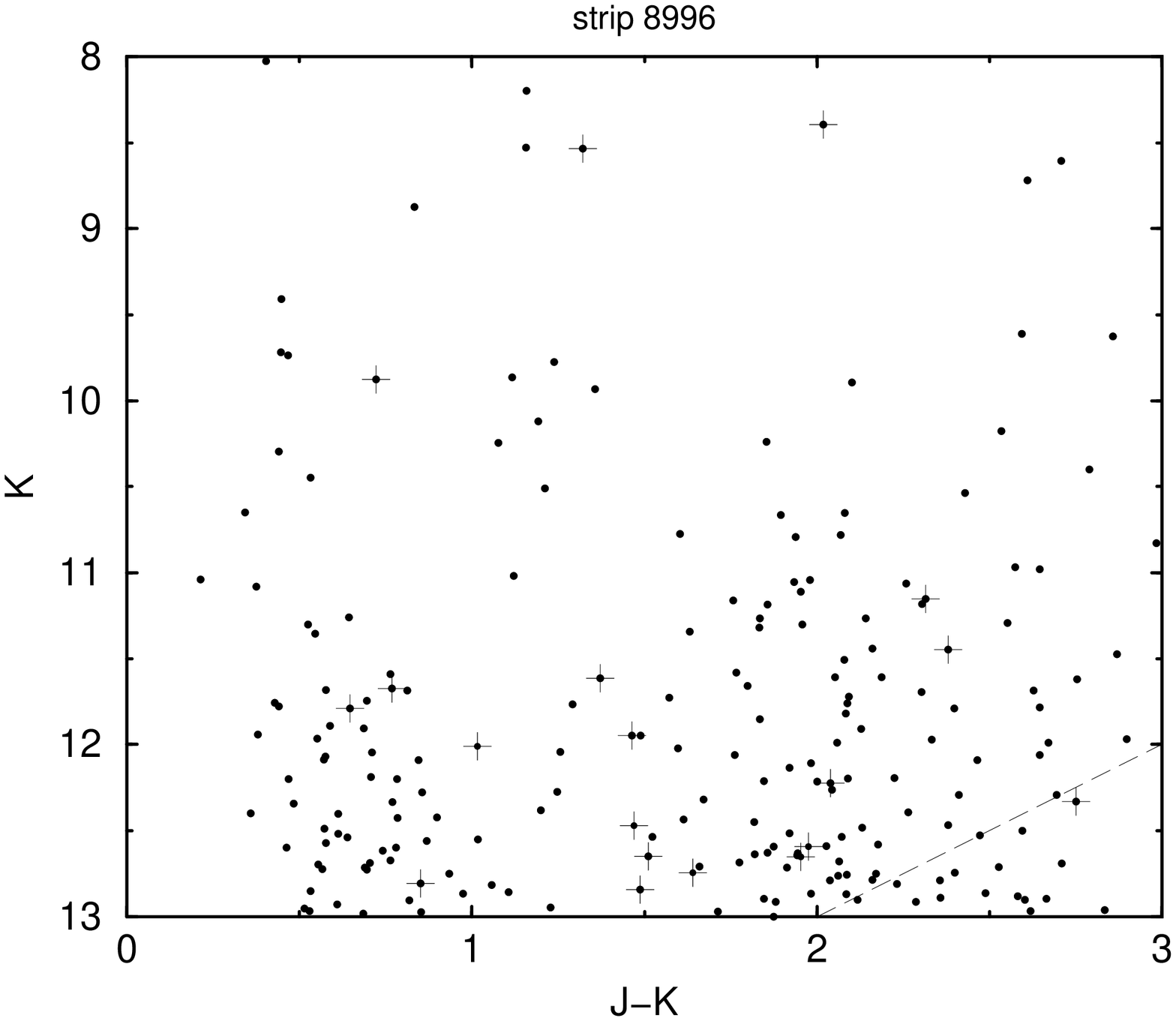}\hspace*{0.01cm}
\includegraphics[width=5.1cm,clip=,angle=-90,bb=97.5 111 572 637]{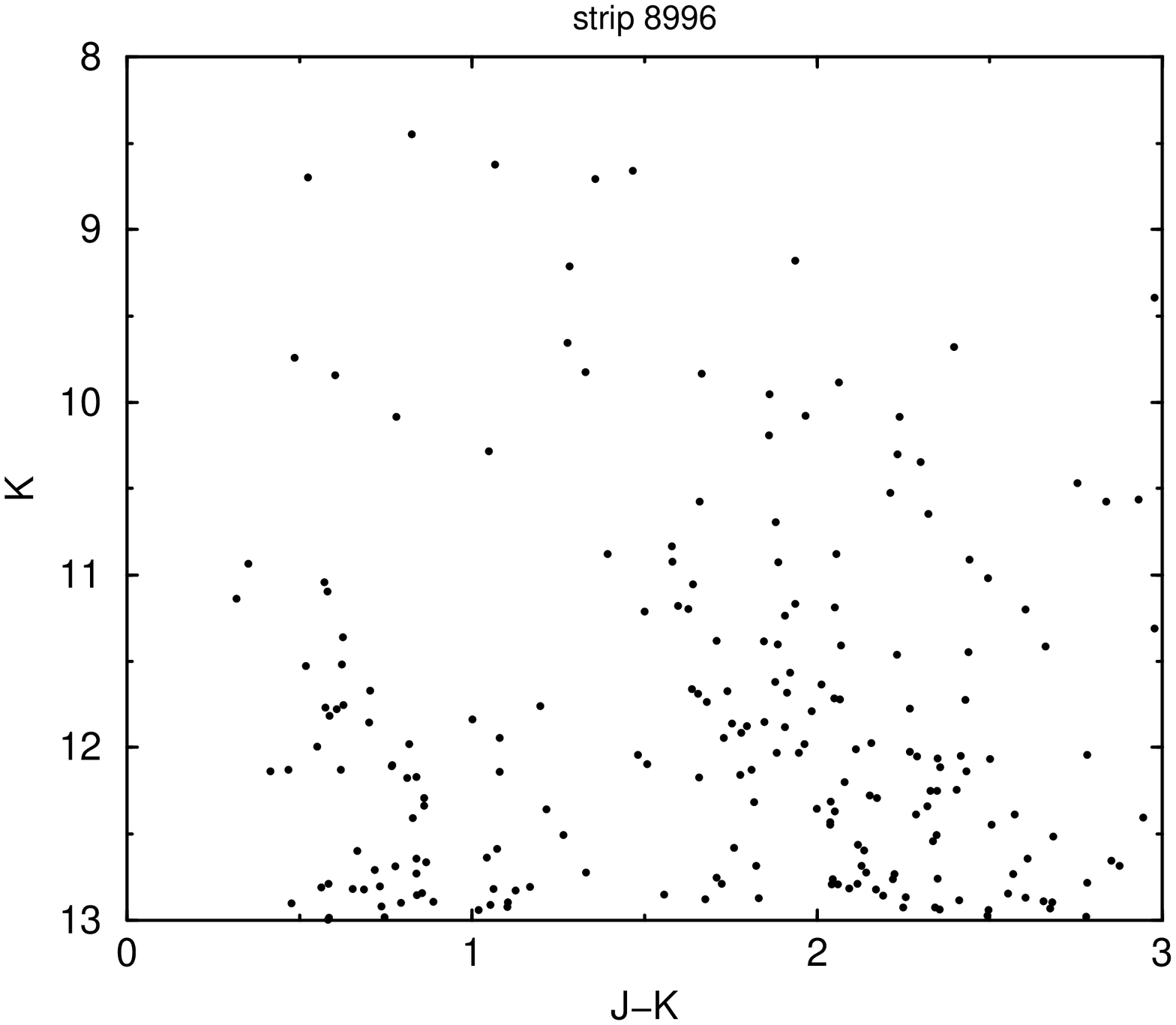}\hspace*{0.01cm}
\includegraphics[width=5.1cm,clip=,angle=-90,bb=97.5 111 572 637]{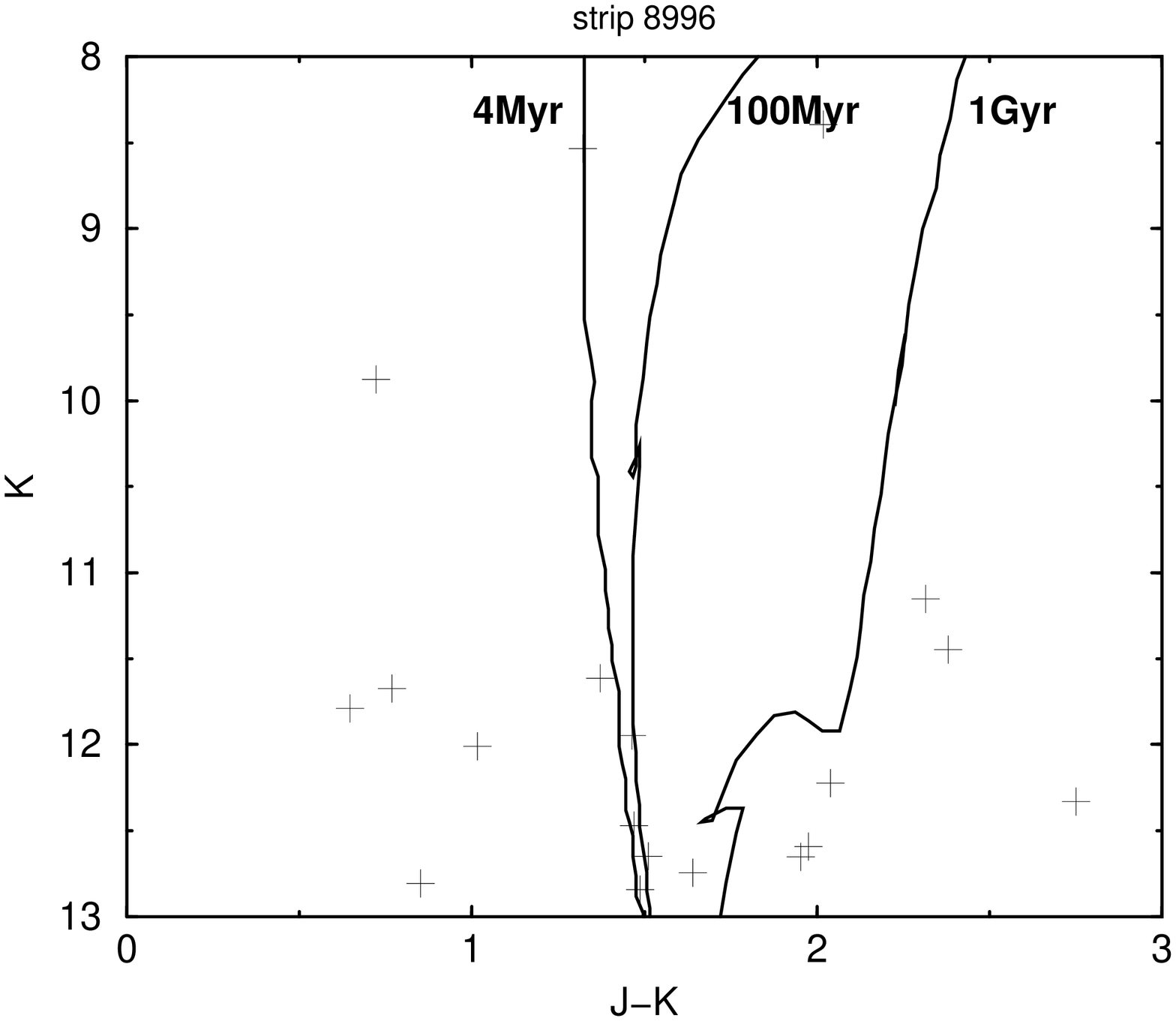}\hspace*{0.01cm}
   \caption{Color-magnitude diagram (K$_s$,J-K$_s$) for Gum 25 cluster. a: Observed
diagram. Field stars are represented with dots, cluster stars with pluses.
Photometric errors are 0.05 at K$_s$ = 12 
and 0.15 at K$_s$ = 13. The dashed line 
shows the J completeness limit. b: Simulated diagram (see text). c: Observed 
diagram
for cluster stars 
only, superimposed with Padova isochrones of different ages.}
   \label{fig2}
\end{figure*}

\begin{figure*}
\hspace*{.8cm} (a) \hspace*{5.15cm} (b) \hspace*{5.15cm} (c) \vspace*{-.75cm} \\
\includegraphics[width=5.1cm,clip=,angle=-90,bb=97.5 60 572 637]{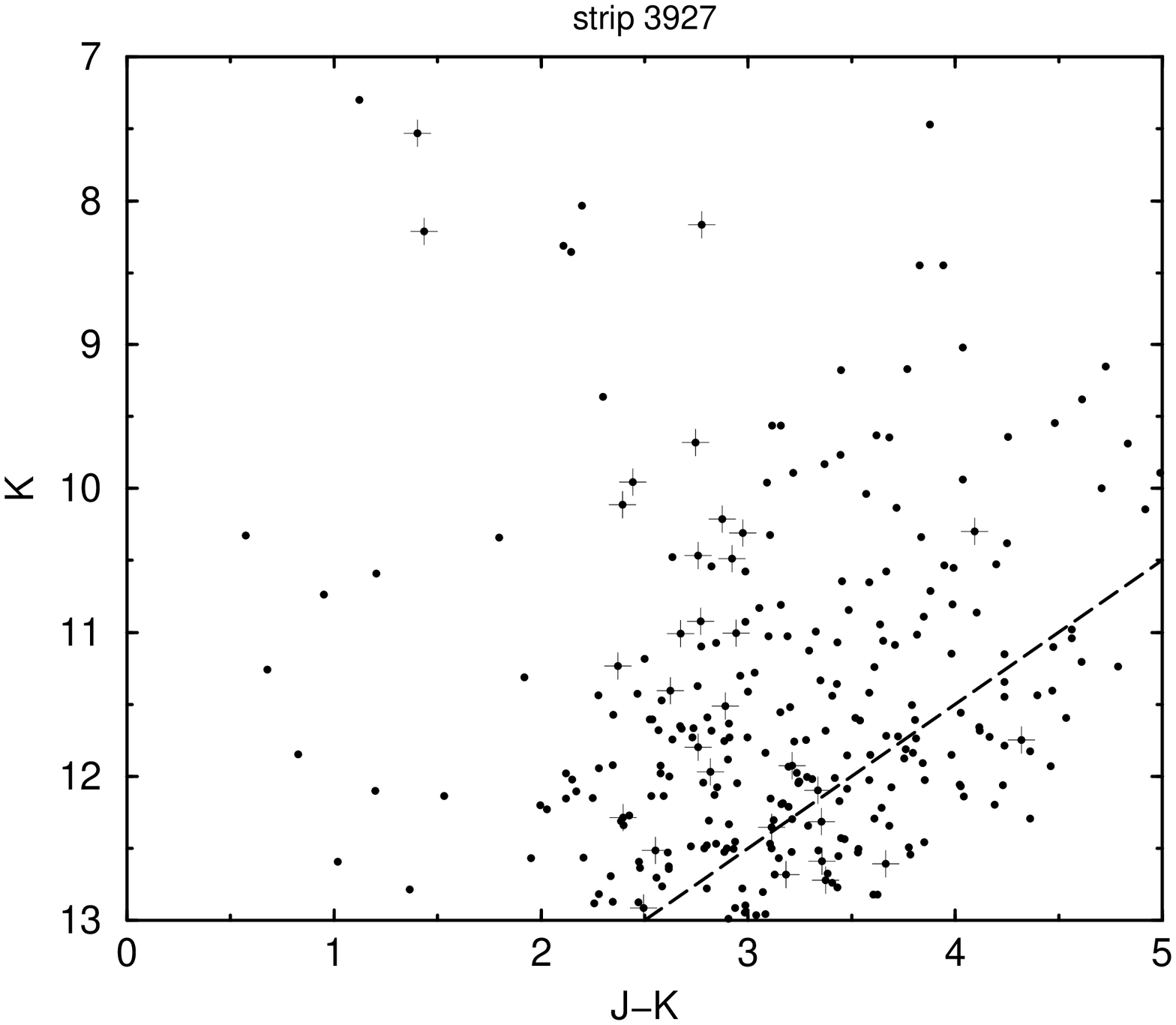}\hspace*{0.01cm}
\includegraphics[width=5.1cm,clip=,angle=-90,bb=97.5 111 572 637]{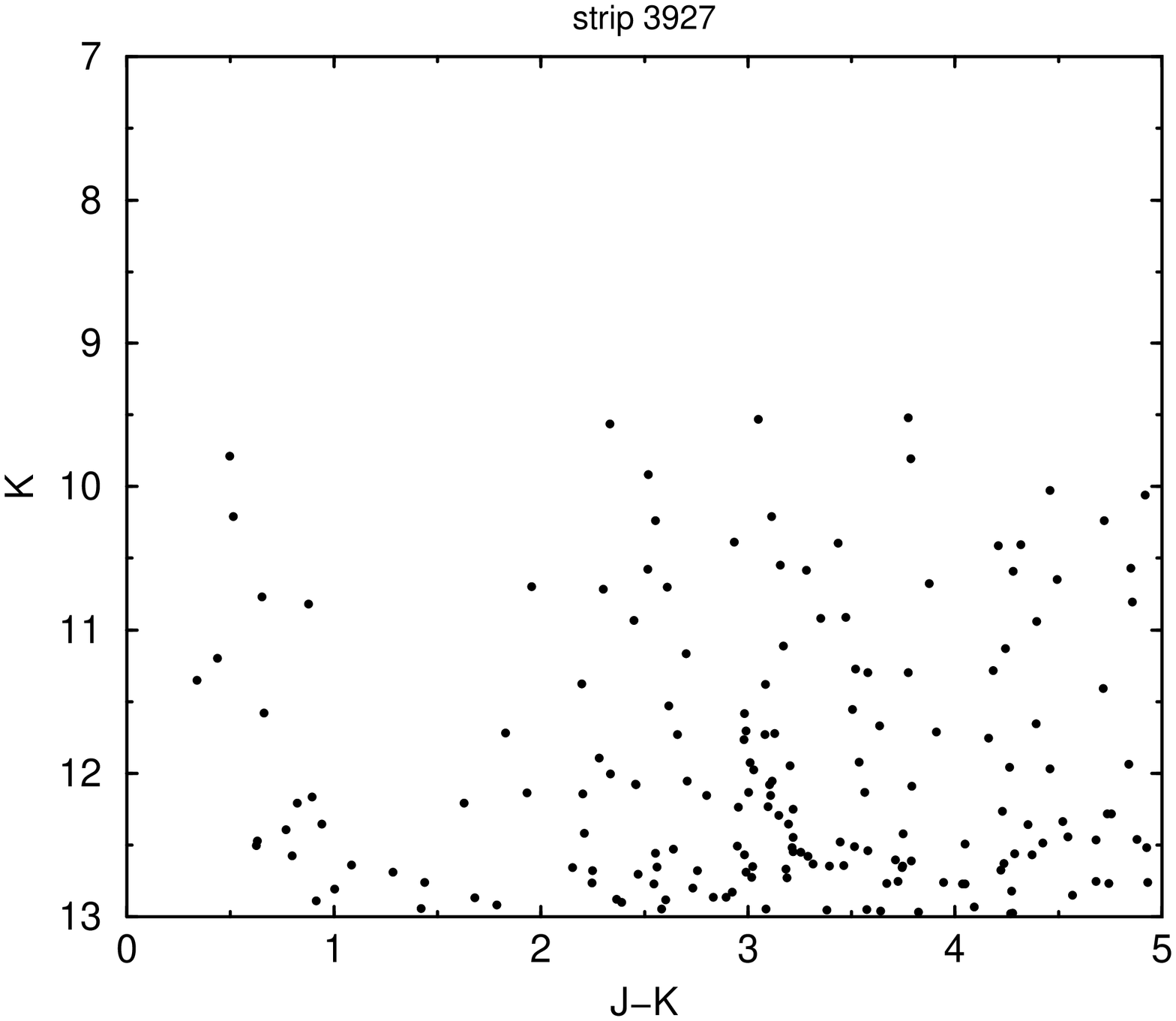}\hspace*{0.01cm}
\includegraphics[width=5.1cm,clip=,angle=-90,bb=97.5 111 572 637]{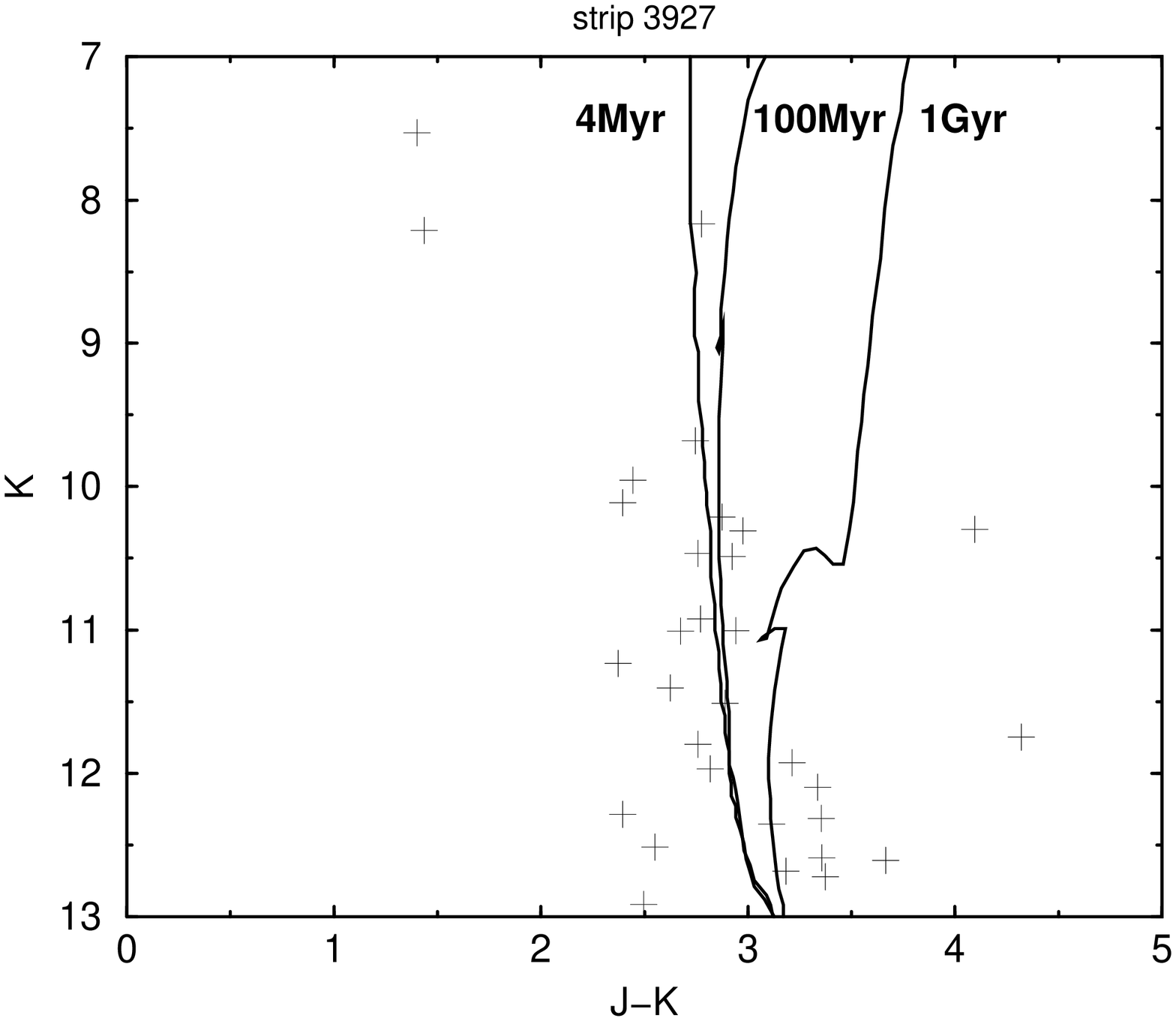}\hspace*{0.01cm}
   \caption{Same as Fig.~\ref{fig2} for W40 cluster. Photometric errors are 
0.2 at K$_s$ = 11 and 0.3 at K$_s$ = 13.}
   \label{fig3}
\end{figure*}

Already known star clusters in the studied region detected with this method 
are listed in Table~1. Most of the non-detected clusters are faint and 
do not appear clearly in the DENIS images.  NGC 6544 and 
NGC 6553 were not detected whereas they are bright globular clusters. The 
first one is effectively not revealed in the catalogue probably because the 
bright part in the center has been extracted as a single bright source. The 
second one is in between two strips and only partially visible in one 
strip, whereas the adjacent strip is not available. Two star 
clusters not yet published have been identified. However, these clusters have 
been previously
presented in the ``Picture of the week'' 2MASS web site, as embedded clusters
in HII regions Gum 25 and W40.
Properties of these clusters, hereafter Gum 25 cluster and W40 
cluster, are given in Table~2. Images in the K$_s$-band are shown 
in Fig.~\ref{fig1}. 
(K$_s$,J-K$_s$) color-magnitude diagrams are plotted in 
Figs.~\ref{fig2}a and \ref{fig3}a for Gum 25 cluster and W40 cluster 
respectively. Field stars are plotted with dots, cluster stars 
with pluses. 

\begin{table}[h]
\caption{Known clusters in the studied region detected with the method 
described in Sect.~\ref{method}. *C: star clusters. OC: open clusters. GC:
globular clusters.}
\begin{tabular}{p{2.1cm}p{2.8cm}p{0.8cm}p{0.7cm}p{0.3cm}}
\hline
name           &\multicolumn{1}{c}{$\alpha$ \hspace*{0.1cm} (J2000)\hspace*{0.1cm} $\delta$}  &\multicolumn{1}{c}{l($^o$)} &\multicolumn{1}{c}{b($^o$)} &\\
\hline
LDN1654 $^a$   &06 59 42 \hspace*{0.1cm} $-$07 46 29 &220.79 &$-$1.71 &*C\\
Pismis 2       &08 17 55 \hspace*{0.1cm} $-$41 40 18 &258.86 &$-$3.34 &*C\\
NGC 2660       &08 42 18 \hspace*{0.1cm} $-$47 09 00 &265.85 &$-$3.03 &OC\\
Pismis 12      &09 19 54 \hspace*{0.1cm} $-$45 08 00 &268.64 &+3.19 &*C\\
Westerlund 2   &10 24 02 \hspace*{0.1cm} $-$57 45 30 &284.27 &$-$0.33 &OC\\
NGC 4337       &12 23 54 \hspace*{0.1cm} $-$58 08 00 &299.29 &+4.54 &*C\\
NGC 5927       &15 28 44 \hspace*{0.1cm} $-$50 40 22 &326.70 &+4.79 &GC\\
NGC 5946       &15 35 29 \hspace*{0.1cm} $-$50 39 35 &327.58 &+4.19 &GC\\
Lyng\aa~7      &16 11 03 \hspace*{0.1cm} $-$55 18 52 &328.77 &$-$2.79 &GC\\ 
NGC 6256       &16 59 33 \hspace*{0.1cm} $-$37 07 17 &347.79 &+3.30 &GC\\
Tonantzintla 2 &17 36 11 \hspace*{0.1cm} $-$38 33 13 &350.80 &$-$3.42 &GC\\
Terzan 5       &17 48 05 \hspace*{0.1cm} $-$24 46 48 &3.84   &+1.69 &GC\\
NGC 6440       &17 48 53 \hspace*{0.1cm} $-$20 21 34 &7.73   &+3.80 &GC\\ 
NGC 6441       &17 50 13 \hspace*{0.1cm} $-$37 03 04 &353.53 &$-$5.01 &GC\\
2MASS GC01     &18 08 22 \hspace*{0.1cm} $-$19 49 47 &10.47  &+0.10 &GC\\
2MASS GC02     &18 09 36 \hspace*{0.1cm} $-$20 26 44 &10.07  &$-$0.45 &GC\\
Terzan 12 $^b$ &18 12 16 \hspace*{0.1cm} $-$22 44 31 &8.36   &$-$2.10 &GC\\
NGC 6712       &18 53 04 \hspace*{0.1cm} $-$08 42 21 &25.35  &$-$4.30 &GC\\
NGC 6749       &19 05 15 \hspace*{0.1cm} +01 54 03 &36.20  &$-$2.20 &GC\\
NGC 6760       &19 11 12 \hspace*{0.1cm} +01 01 50 &36.10  &$-$3.92 &GC\\
NGC 6530       &18 04 48 \hspace*{0.1cm} $-$24 20 00 &6.14   &$-$1.38 &*C\\
\hline
\end{tabular}
$a$ \cite{Hodapp1994}\\
$b$ also designed as ESO522SC1
\end{table}

\begin{table}[h]
\caption{Gum 25 cluster and W40 cluster characteristics. Both are 
embedded in indicated HII regions.}
\begin{tabular}{lll}
\hline
                                  &Gum 25 cluster 	 &W40 cluster \\
\hline 
$\alpha$,$\delta$ (J2000)         &09 02 11,$-$48 49 14  &18 31 25,$-$02 05 02\\
$l,b$                             &269.27,$-$1.53        &28.80,+3.50\\
size                              &1.4$' \times 1.2''$    &1.9$' \times 1.3'$\\
$A_V$	                          &9 mag		 &17 mag\\
HII region                        &Gum 25 $^a$	         &W40 $^b$\\
\hspace{0.1cm} --- distance $^c$  &1.8 kpc               &600 pc\\
\hspace{0.1cm} --- $V_{\mbox{\scriptsize LSR}}$ $^d$ &7 km s$^{-1}$         &0.6 km s$^{-1}$\\
\hline
\end{tabular}
$a$ also designed as RCW40\\
$b$ also designed as Sh2-64, RCW174,[L89b]28.790+0.346\\
$c$ Gum 25: \cite{Avedisova1989}, W40: \cite{Wu1996}\\
$d$ Gum 25: \cite{Brand1987}, W40: \cite{Lockman1989}
\end{table}

\subsection{Determination of cluster characteristics}

While the DENIS data alone do not allow to determine the definitive membership
of a star to a cluster (kinematics or distances would be needed), one may estimate
a probable membership from the position on the sky and estimate the 
contamination by field stars. From the density around the clusters, the 
expected contamination in both clusters is 5 and 6 field stars respectively. 
The cluster size is estimated visually from the image and catalogue, 
as well as the stellar content. We have attempted to estimate the reddening
of the clusters and in the field by fitting an extinction variable along
the line of sight. The Besan\c{c}on model of population synthesis 
\citep{Haywood1997} is used to 
reproduce the stellar content in the fields. The extinction on the
pathsight is fitted in order to reproduce the stellar distribution in the
(K$_s$,J-K$_s$) plane. It is done using a distribution of diffuse extinction 
which
is produced by a double-exponential disc of interstellar material, plus 
discrete clouds as needed on the line of sight. The overall extinction 
at position ($l,b$) and distance $d$ writes:
\[A_V(l,b,d) = \int_0^d A_{\mbox{\scriptsize dif}} \times e^{-\frac{R}{hl}}\times 
e^{-\frac{z}{hh}} dr + \sum Cl(l,b,d)\]
where $hl$ and $hh$ are the extinction scale length and scale height, 
$Cl(l,b,d)$ the extinction due to the intervening cloud in $(l,b)$ direction 
at distance $d$, $(R,z)$ the galactic coordinates. The scale length and scale 
height are taken as 4.5 kpc and 140 pc, following the expected distribution 
of 
the dust. $A_{\mbox{\scriptsize dif}}$ is the differential diffuse
extinction at the solar position in magnitude per kpc. This value is often
taken between 0.5 and 1. This range of value works well at high and 
intermediate latitudes but often failed in the plane. It so need to be 
adjusted on each line of sight.

The distribution of extinction described above gives a univoque relation
between distance and extinction in a given field. However observations
sometimes show rapid variations of extinction from one line of sight to 
another especially in dense clouds. This can be accounted for by assuming 
a dispersion of extinction
for any given star (each star being on a different line of sight). 
\citet{Lada1994} found in the cloud IC5146 variations of extinction which can 
be modeled with dispersion proportional to the extinction: 
$\sigma = 0.73 + 0.40 \times A_V$ in a field of 1.5' on a side. 
\citet{Thoraval1997} find a lower value: $\sigma = 0.25 \times A_V $.
In adjusting our model of extinction we have tentatively used either a
constant extinction at a given distance or variations proportional to the
extinction with the above formula.

\subsection{Gum 25 cluster}

In this field it has been easy to describe the extinction with a simple
model of diffuse extinction distributed over the line of sight with no
high density cloud. The (K$_s$,J-K$_s$) diagram of field stars is well 
reproduced with a differential diffuse extinction at the solar position
$A_{\mbox{\scriptsize dif}}$ = 2.5 mag kpc$^{-1}$ (Fig.~\ref{fig2}b). 
With this model, the extinction reaches about 10 magnitudes at 10 kpc.

Using theoretical Padova isochrones \citep{Bertelli1994,Girardi1996}, we
try to estimate the reddening of the cluster. However
it appears difficult to fit a well defined isochrone on the color-magnitude
diagram of this cluster due to the large dispersion in colour. 
The age cannot be constrained either.
This dispersion 
cannot be interpreted by field star contamination alone and must be due to a 
strongly varying extinction inside or in front of the cluster. 
We estimate the mean extinction to be $A_V$=9 mag.
Fig.~\ref{fig2}c shows the cluster (K$_s$,J-K$_s$) diagram 
together 
with the isochrones at a distance of 1.8 kpc, which is the distance of the 
Gum 25
HII region, and with a visual extinction of 9 magnitude. 
This is higher than the value proposed by 2MASS on their web page, $A_V$=3.6 mag,
a value which we propose as a lower limit for less extincted stars in the 
cluster.


\subsection{W40 cluster}

The distribution of extinction in this field has been more difficult
to estimate. The simple extinction model from the interstellar matter disc
does not allow to explain why there are so few main sequence stars in this
field. A high extinction cloud at 600 pc from us, which is the distance of 
the W40 ionized region, allows however to 
reproduce the number of main sequence stars if this cloud has a visual 
extinction of 11 magnitudes. In this case the other background stars are 
reddened to J-K $>$ 2. The (K$_s$,J-K$_s$) diagram of simulated 
stars is given in Fig.~\ref{fig3}b. Main sequence stars are at distances 
less than 2 kpc, while giants appear at 4-8 kpc. At a distance of 6 kpc
in the direction of W40 cluster, the stars are above the diffuse extinction 
disc.
Thus the diffuse extinction is not sufficient to explain the large dispersion 
among field giants. In order to reproduce the distribution in J-K$_s$ of the 
giants  as well as the observed dispersion, we adopt
an internal dispersion on the extinction at a given distance of $\sigma = 
0.25 \times A_V$. The distribution of extinction along the line of sight is 
defined by $A_{\mbox{\scriptsize dif}}$ = 3 mag kpc$^{-1}$, and 11 mag 
extinction due to a cloud at 600~pc. With this model, the extinction reaches 
about 21 magnitudes at 10 kpc.

In the cluster, the isochrone fitting of the main sequence is obtained with 
the cluster being at the same distance as the main cloud d = 600~pc and an 
extinction in the cluster $A_V$ = 17 mag, 
larger than the cloud which reddens the field. It can be interpreted as 
the central dense part of a larger cloud. Fig.~\ref{fig3}c shows the main 
sequence fitting of the cluster with Padova isochrones of 4~Myr, 100~Myr and 
1~Gyr, visual extinction of 17 mag and distance of 600~pc. It shows that the 
cluster is rather young.

\section{Conclusion}
We elaborated a program to perform an automated search of star clusters from
the catalogues of extracted sources from the DENIS survey. We concentrated
on the band around the Galactic plane ($|b| < 5^o$), where 44\% of
the data have already been calibrated. Most of the already known star clusters
visible on the DENIS images in the probed region have been detected. Two
star clusters, not published yet, have been identified. Both clusters are
embedded in their associated HII regions. Extinction in front of the clusters is
estimated. Uncertainties remain large on this
determination mainly due to questionnable membership and to the photometric errors at
faint magnitudes. Deeper and more accurate photometry and proper motions
would allow to assert the cluster characteristics.

Although we have missed known star clusters and probably not yet detected
clusters, our method applied to Near-Infrared data
allows to search for old star clusters and embedded star clusters in a systematic 
way, with a lower efficiency
than when looking at the image, but much more rapidly.
We plan to run again this program once the complete region around the
Galactic plane will be calibrated.

\begin{acknowledgements}
The authors thank Cyril Falvo who helped in the programming, the whole DENIS 
staff and all the DENIS observers who collected the data. The DENIS project is 
supported by the SCIENCE and the Human Capital and Mobility plans of the 
European Commission under grants CT920791 and CT940627 in France, by 
l'Institut National des Sciences de l'Univers, the Minist\`ere de 
l'\'education Nationale and the Centre National de la Recherche Scientifique 
(CNRS) in France, by the State of Baden-W\"urtemberg in Germany, by the 
DGICYT in Spain, by the Sterrewacht Leiden in Holland, by the Consiglio 
Nazionale delle Ricerche (CNR) in Italy, by the Fonds zur F\"orderung der 
wissenschaftlichen Forschung and Bundesministerium f\"ur Wissenschaft und 
Forsch ung in Austria, and by the ESO C \& EE grant A-04-046.
\end{acknowledgements}

\end{document}